\documentclass[a4paper,10pt]{article}
\usepackage[a4paper, margin=2cm]{geometry}

\usepackage[utf8]{inputenc}
\usepackage[T1]{fontenc}

\usepackage{fourier}

\usepackage[square,numbers]{natbib}
\usepackage{amsmath}
\allowdisplaybreaks
\usepackage{amsfonts}
\usepackage{bm}
\usepackage{mathrsfs}
\usepackage{booktabs}
\usepackage{graphicx}
\usepackage{url}
\usepackage{xcolor}

\usepackage{amsthm}
\theoremstyle{remark}

\usepackage[english]{babel}
\usepackage[hidelinks]{hyperref}

\usepackage{doi}

\newcommand\blfootnote[1]{%
  \begingroup
  \renewcommand\thefootnote{}\footnote{#1}%
  \addtocounter{footnote}{-1}%
  \endgroup
}

\title{The KP equation of plane elastodynamics}
\author{Harold Berjamin \textsuperscript{a}, Michel Destrade \textsuperscript{a}, Giuseppe Saccomandi \textsuperscript{b,a} \\ ~ \\
\emph{\footnotesize\textsuperscript{a}School of Mathematical and Statistical Sciences, University of Galway, University Road, Galway, Republic of Ireland} \\
\emph{\footnotesize\textsuperscript{b}Dipartimento di Ingegneria, Universit\`a degli Studi di Perugia, 06125 Perugia, Italy} }

\date{}

\renewcommand{\vec}[1]{\boldsymbol{#1}}
\begin{document}

	\maketitle
	
	\begin{abstract}
		\noindent
		The propagation of nonlinear and dispersive waves in various materials can be described by the well-known Kadomtsev–Petviashvili (KP) equation, which is a (2+1)-dimensional partial differential equation. In this paper, we show that the KP equation can be used to describe the in-plane motion of compressible elastic solids with dispersion. Furthermore, a modified KP equation with cubic nonlinearity is obtained in the case of incompressible solids with dispersion. Then, several solutions of these partial differential equations are discussed and computed using a Fourier spectral method. In particular, both equations admit solitary wave solutions.
		~ 
		\blfootnote{{\bfseries Funding.} HB has received funding from the European Union’s Horizon 2020 research
and innovation programme under grant agreement TBI-WAVES – H2020-MSCA-IF-2020 project No 101023950. GS is partially supported by the
PRIN research project ‘The mathematics and mechanics of nonlinear wave
propagation in solids’ (grant no. 2022P5R22A), and by GNFM of Istituto
Nazionale di Alta Matematica (INDAM).}
	\end{abstract}

\section{Introduction}
The asymptotic limits of many nonlinear wave phenomena are universally and canonically described by a small number of \emph{model equations}, such as the  Burgers equation, the Korteweg--De Vries (KdV) equation or the nonlinear Schrödinger equation (NLSE) \cite{DP}. 
These models provide a common playground to study nonlinear waves in a disparate ensemble of widely different physical systems.

It is not entirely surprising that only a few canonical and universal asymptotics equations occur over and over again. 
For example, in Classical Mechanics, studying small oscillations about a stable equilibrium is an asymptotic method that works for all Lagrangian systems and always provides a second-order linear system of differential equations. 
Hence, for any discrete Lagrangian system we obtain the ``same'' set of differential equations, up to different numerical values in the mass and stiffness matrices \cite{Biscari}.   

In general, model equations exist in a nonlinear framework and their derivation requires more sophisticated perturbative methods than a regular perturbation scheme. 
Hence, to provide a uniformly valid approximation for small-am\-pli\-tu\-de nonlinear waves over long times and long spaces, we have to turn to the method of multiple scales \cite{Taniuti}.

In Fluid Mechanics, the rigorous derivation of model equations from basic principles is a well-established field, and it is presented and analysed in many textbooks and survey articles, such as for example those by Crighton \cite{Cri} or Johnson \cite{J}. 
In Solid Mechanics, the situation is not, as yet, as well defined. 
Insightful information can be obtained from the book \cite{Mauginbook} and survey \cite{Maugin} by Maugin as well as the monographs by Samsonov \cite{Samsonov} and Porubov \cite{Porubov}, but some big pieces of the puzzle are still missing. 

One example of model equation is the classical Kadomtsev–Petviashvili (KP) equation \cite{KP70}, a (2+1)-dimensional (two space variables, one time variable) version of the KdV equation, characterised by a very rich structure, and which might be considered as the most fundamental integrable system. 
In standard form it reads \cite{Hirota, deBouard, Pouget03}
\begin{equation}
(U_t + 6 UU_x + U_{xxx})_x = \pm U_{yy},
\label{KP-standard}
\end{equation}
where the sign on the right-hand side is determined by several factors, some of which we will identify in our derivation below. 
Conventionally, the plus and minus signs correspond to the the so-called KP-I and KP-II equations, respectively. 

The status of this equation is well recognised in Fluid Mechanics (see for example \cite{Kodama}), but not so much in Solid Mechanics. 
It seems that the emergence of this equation in the framework of a nonlinear theory related to solid mechanics can be traced back to a handful of papers only. 

Maugin \cite{Mauginbook}  claims that a derivation of the KP equation may be found in a 1991 paper by Pouget \cite{Pouget}.
Upon inspection, however, there we find an asymptotic equation which differs from the KP model equation both in the nature of the nonlinearity and in the higher-order derivatives involved. 
Indeed, its nonlinear terms are those of the Gardner equation \cite{Miura}, while the third-order derivatives are of  ``Zakharov-Kuznetsov type modelling'', according to Pouget.  
In a subsequent paper, Pouget \cite{Pouget95} derives a KP equation with second/third-order mixed nonlinearity. 

Maugin also references two papers published in the proceedings  of the eighth international symposium on \emph{Continuum Models and Discrete Systems}  \cite{Markov}, which took place in Bulgaria in 1995. 
The first paper, by Collet and Pouget, is about a thin plate resting on elastic foundations, with an asymptotic procedure leading, not to a KP equation, but to a (2+1)-dimensional nonlinear Schroedinger equation. 
The second paper, by Erbay, presents a system of coupled modified KP equations in the framework of micropolar elasticity \cite{Erbay}. 
From this brief survey we conclude that a simple and clear derivation of the classical KP equation in the framework of isotropic nonlinear elasticity seems to be missing.

% The outline is not required, but we show an example here.
In the next section (Section \ref{section2}) we use the fundamental asymptotic scheme proposed by Zabolotskaya \cite{ZZ} to obtain a two-dimensional asymptotic equation from the general equations of motion in plane isotropic elasticity. 
As expected, this equation is similar to the one obtained by Zabolotskaya \cite{ZZ}, but our derivation is different, and to some extent, more general. Here, we derive the equation for an arbitrary strain-energy density function and then specialise it to that of third-order weakly nonlinear elasticity. In contrast, Zabolotskaya starts from the governing equations of third-order elasticity to arrive at the asymptotic equation. 
We then introduce a model of dispersive elasticity \cite{Destradecompact} to arrive at a two-dimensional Burgers equation and a KP equation.

A similar approach was followed in Reference~\cite{DestradeGoriely} for the study of antiplane shearing motions in incompressible solids. The main difference with the present study is the type of 2D motion that is considered. In the present case of plane elasticity, the motion of particles is restricted to a given plane of interest. In antiplane shear, there is no in-plane particle motion; instead, particles are only allowed to move along the direction perpendicular to the plane of interest. Based on suitable asymptotic expansions, a modified KP equation with cubic nonlinearity is obtained in this case.

In Section \ref{section3}, we discuss the incompressible case.
Although it can be derived from the equations in the compressible case as a limit, it brings out significant differences in the treatment. 
Hence the asymptotic scalings must be adjusted and the resulting KP equation has \emph{cubic} nonlinearity. 
This is an interesting development, because it is rare to see the cubic KP equation emerge from the equations of elastodynamics: 
it seems that the only instances are works by Pouget \cite{Pouget} using the continuum limit of a lattice model, and by Erbay \cite{Erbay} and Babaoglu and Erbay  \cite{Babaoglu}, using micromorphic elasticity.
Here we derive it from the equations of nonlinear elasticity (incorporating dispersion), with no assumptions made on the microstructure of the continuum\,---\,see also Ref.~\cite{DestradeGoriely} for the case of antiplane shear. Note in passing that the KP equation with cubic nonlinearity is sometimes referred to as the modified KP equation (mKP) in other works \cite{DK}.
 
Section \ref{section4} is devoted to the study of some particular solutions to the quadratic and cubic KP equations. In particular, computational results were obtained by using a Fourier spectral method. In the final section, we present concluding remarks. An alternative derivation of the results obtained in the incompressible case (Section \ref{section3}) is provided in the Appendix \ref{StreamFunc}.

%%%%%%%%%%%%%%%%%%%%%%%%%%%%%%

\section{Plane motions in compressible solids}
\label{section2}

%%%%%%%%%%%%%%%%%%%%%%%%%%%%%%%

\subsection{Governing equations}

We start with the classical general theory of isotropic hyperelasticity. 
In this framework the strain-energy density, $W$, is a function of three scalars, $W=W(I_1, I_2, J)$, where 
\begin{equation}
I_1=\text{tr} \, \vec{F}^T\vec F, \qquad J = \det \vec{F}, \qquad I_2 = J^2 \,\text{tr}(\vec{F}^{-1}\vec{F}^{-T}), 
\end{equation}
and $\vec{F}$ is the deformation gradient. 
 
We restrict our attention to the family of \emph{plane motions}
\begin{equation} \label{e23}
x=X+u(X,Y,t), \quad  y=Y+v(X,Y,t), \quad z=Z,
\end{equation}
where $x,y,z$ and $X,Y,Z$ are the Cartesian coordinates in the current and reference configurations, and $u$ and $v$ are the in-plane displacements.
We then find 
\begin{equation}
\vec{F} = \begin{bmatrix}
1+u_X & u_Y & 0\\
v_X & 1+v_Y & 0\\
0 & 0 & 1
\end{bmatrix},  \qquad 
J\vec{F}^{-1} = 
\begin{bmatrix}
1+v_Y & -u_Y & 0 \\
-v_X & 1+u_X & 0 \\
0 & 0 & J
\end{bmatrix},
\end{equation}
where the subscripts denote partial differentiation. 
Then we readily compute the quantities
\begin{equation}
I_1=1+v_X^2+u_Y^2+(1+u_X)^2+(1+v_Y)^2, \quad 
J=1+u_X+v_Y+u_X v_Y-u_Y v_X.
\label{I1-J}
\end{equation}
We also find that $I_2 = I_1 + J^2 - 1$, so that there are only two independent invariants, say $I_1$ and $J$.

For the in-plane components of the elastic nominal stress tensor $\vec S^e$, we may use the following representation formula \cite{Tarantino}
\begin{equation} \label{stress}
[\vec{S}^e]_{ij} = J \alpha [\vec{F}^{-1}]_{ij} + \gamma [\vec{F}^T]_{ij},
\end{equation}
where the subscripts $i$ and $j$ run over $1$ and $2$ and the constitutive coefficients $\alpha$, $\gamma$ are defined as
\begin{equation} \label{beta-gamma}
\alpha = 2 J \left(\dfrac{\partial W}{\partial I_2} + \dfrac{\partial W}{\partial I_3}\right),
 \qquad 
 \gamma = 2 \left(\dfrac{\partial W}{\partial I_1} + \dfrac{\partial W}{\partial I_2} \right) ,
\end{equation}
where $I_3 = J^2$.
We also have $S^e_{13} = S^e_{23}=0$, and $S^e_{33} = S^e_{33}(X,Y,t)$ only.

Now, the first two lines of the equations of motion, $\text{Div}\,\vec{S}^e = \rho_0 \vec{x}_{tt}$, read
\begin{equation} \label{eqzero}
\begin{aligned}
& \left[\alpha (1+v_Y)+\gamma_{}(1+u_X) \right]_X+ \left[- \alpha v_X + \gamma_{}u_Y \right]_Y=\rho_0  u_{tt},  \\[4pt]
& \left[- \alpha u_Y+\gamma_{} v_X \right]_X+ \left[\alpha(1+u_X)+\gamma_{}(1+v_Y) \right]_Y=\rho_0   v_{tt},
\end{aligned}
\end{equation}
where $\rho_0$ is the mass density in the reference configuration, and the third line reduces to $\partial S^e_{33}/\partial Z = 0$, which is an identity because $S_{33}^e$ is independent of $Z$.
We expand the two equations as
\begin{subequations}\label{eq}
\begin{align} 
& \alpha_{X} (1 + v_Y) - \alpha_{Y} v_X + \gamma_{}(u_{XX} + u_{YY}) + \gamma_{X}(1+u_X) + \gamma_{Y}u_Y=\rho_0  u_{tt},  \label{eqa}\\[4pt]
& {-} \alpha_{X} u_Y + \alpha_{Y}(1 + u_X) + \gamma_{} (v_{XX} + v_{YY}) + \gamma_{X}v_X + \gamma_{Y}(1+v_Y) = \rho_0   v_{tt}. \label{eqb}
\end{align}
\end{subequations}

\subsection{Asymptotic analysis}
The full system \eqref{eq} is  clearly too complex to be solved analytically.
One way to make progress is to introduce a small parameter $\epsilon$ and to implement a perturbative scheme. 
This approach may give some insight on the nonlinear phenomena associated with the equations but, as already pointed out, only under the assumption that the perturbation is {small} and that  the propagation distance and time are  {not too large}. 
To provide a uniformly valid approximation for small amplitude nonlinear waves over long times and long spaces we have to use the method of multiple scales. 

For  nonlinear elastic waves, this means that we have to introduce new, dimensionless, amplitude functions $ \tilde{u}$,  $\tilde v$, such that 
\begin{equation} \label{ampl}
u=\epsilon L \tilde{u}, \qquad v=  \epsilon^{3/2} L \tilde{v},
\end{equation}
where $L$ is a characteristic length, and new, dimensionless, space and time variables
\begin{equation} \label{space}
\chi=\epsilon \dfrac{X}{L}, \qquad \eta=\sqrt{\epsilon} \dfrac{Y}{L}, \qquad \tau= \dfrac{1}{L}(ct - X),
\end{equation}
where $c$ is a constant with the dimensions of a speed.  
The time variable $\tau$ describes the evolution of the particle displacement in the moving frame following the wave during its propagation in the $X$ direction at the speed $c$. We note that the leading-order displacement component is that along the $X$-axis, $u$, which is of order $\mathcal{O}(\epsilon)$. Furthermore, its leading-order gradient is that along the $X$-axis, $u_X$, see the calculations below. Thus, 
this scaling gives a \emph{quasi} one-dimensional system because the signal is localised in the direction transverse to the direction of wave propagation, in line with what we observe in many nonlinear wave propagation phenomena.

We  stop expansions at order $\mathcal{O}(\epsilon^2)$.  
First, we find
\begin{align}
& u_X = - \epsilon \tilde u_\tau + \epsilon^2 \tilde u_\chi,
 && u_Y = \epsilon^{3/2} \tilde u_\eta,
&&  u_{XX} + u_{YY} = \frac{1}{L}(\epsilon \tilde u_{\tau\tau} - 2 \epsilon^2 \tilde u_{\chi\tau} + \epsilon^2 \tilde u_{\eta\eta}), \notag \\[4pt]
 & v_X = - \epsilon^{3/2} \tilde v_\tau, 
& &  v_Y = \epsilon^{2} \tilde v_\eta,
&& v_{XX} + v_{YY} = \frac{1}{L} \epsilon^{3/2} \tilde v_{\tau\tau}.
\end{align}
and then, 
\begin{equation} \label{J}
I_1-3 =  - 2 \epsilon \tilde{u}_\tau +  \epsilon^2 \left[ 2(\tilde{u}_\chi + \tilde{v}_\eta) +  (\tilde{u}_\tau)^2\right],
\quad
J-1 = - \epsilon\tilde{u}_\tau +  \epsilon^2 (\tilde{u}_\chi + \tilde{v}_\eta).
\end{equation}

Now, recalling that here the strain energy depends on two scalars only, $I_1$ and $J$, say, we may expand the coefficients $\alpha$ and $\gamma$ into Taylor series, as
\begin{equation}
 \alpha = \sum_{i=0}^{2}\sum_{j=0}^2  \alpha_{ij}(I_1-3)^i(J-1)^j, \qquad
 \gamma =  \sum_{i=0}^{2}\sum_{j=0}^2   \gamma_{ij}(I_1-3)^i(J-1)^j, 
 \label{taylor}
\end{equation}
where $\alpha_{ij}$, $\gamma_{ij}$ are constants.
Using the expansions \eqref{J}, we obtain,  up to order $\epsilon^2$,
\begin{equation}
\begin{aligned}
& \alpha = \alpha_{0} - \epsilon \alpha_1 \tilde{u}_\tau + \epsilon^2\left[\alpha_1(\tilde{u}_\chi + \tilde{v}_\eta) + \alpha_2 (\tilde{u}_\tau)^2\right],
\\
& \gamma = \gamma_{0} - \epsilon \gamma_1 \tilde{u}_\tau + \epsilon^2\left[\gamma_1(\tilde{u}_\chi + \tilde{v}_\eta) + \gamma_2 (\tilde{u}_\tau)^2\right],
\end{aligned}
\end{equation}
where 
\begin{equation}
\begin{aligned}
&\alpha_0 = \alpha_{00}, && 
\alpha_1 = 2\alpha_{10} + \alpha_{01},
&&
\alpha_2 = \alpha_{10} + 4 \alpha_{20} + 2 \alpha_{11} + \alpha_{02},
 \\
&\gamma_0 = \gamma_{00}, && 
\gamma_1 = 2\gamma_{10} + \gamma_{01},
&&
\gamma_2 = \gamma_{10} + 4 \gamma_{20} + 2 \gamma_{11} + \gamma_{02},
\end{aligned}
\end{equation}
and then
\begin{equation}
\begin{aligned}
& \alpha_X = \dfrac{\epsilon}{L}\alpha_1\tilde u_{\tau\tau} - \dfrac{\epsilon^2}{L}[\alpha_1(2\tilde u_{\chi\tau}+\tilde v_{\eta\tau}) + \alpha_2\left(\tilde u_\tau^2\right)_\tau], 
&&
\alpha_Y = -\dfrac{\epsilon^{3/2}}{L}\alpha_1\tilde u_{\eta\tau}, 
 \\[4pt]
& \gamma_X = \dfrac{\epsilon}{L}\gamma_1\tilde u_{\tau\tau} - \dfrac{\epsilon^2}{L}[\gamma_1(2\tilde u_{\chi\tau}+\tilde v_{\eta\tau}) + \gamma_2\left(\tilde u_\tau^2\right)_\tau], 
&&
\gamma_Y = -\dfrac{\epsilon^{3/2}}{L}\gamma_1\tilde u_{\eta\tau}.
\end{aligned}
\end{equation}

Now we substitute these expansions in the equations of motion \eqref{eq} and stop  at order $\epsilon^2$. 
Writing \eqref{eqa} at order $\epsilon$ gives
\begin{equation}
(\alpha_1+ \gamma_0 + \gamma_1 - \rho_0 c^2) \tilde{u}_{\tau \tau} = 0,
\label{eq1}
\end{equation}
and at order $\epsilon^2$,
\begin{equation}
 \left[(\alpha_1 +  \gamma_1) \tilde{v}_\eta + 2(\alpha_1 + \gamma_0 + \gamma_1) \tilde{u}_\chi + (\alpha_2 + \gamma_1 + \gamma_2)(\tilde{u}_\tau)^2\right]_\tau - \gamma_{0}\tilde{u}_{\eta \eta} = 0.
\label{eq2}
\end{equation}
And writing \eqref{eqb} up to order $\epsilon^2$ yields a single equation, at order $\epsilon^{3/2}$, which is 
\begin{equation}
(\alpha_1 + \gamma_{1})\tilde{u}_{\eta \tau} + (\rho_0 c^2 - \gamma_{0})\tilde{v}_{\tau \tau}=0.
\label{eq3}
\end{equation}
Taking \eqref{eq1} into consideration, we see that the following identity holds: $\rho_0c^2 - \gamma_0 = \alpha_1 + \gamma_1$, which helps simplify \eqref{eq2} and \eqref{eq3} into 
  \begin{equation} \label{eq0a}
\left[\left(\frac{c_\ell^2}{c_t^2} - 1 \right)\tilde{v}_\eta + 2\frac{c_\ell^2}{c_t^2} \tilde{u}_\chi + 3 \frac{c_\ell^2}{c_t^2}\beta (\tilde{u}_\tau)^2\right]_\tau -  \tilde{u}_{\eta \eta}=0, 
\end{equation}
and
\begin{equation}  \label{eq0b}
\tilde{v}_{\tau \tau} + \tilde{u}_{\eta \tau}=0,
\end{equation} 
respectively, where 
\begin{equation}
c_\ell = c, \qquad
c_t = \sqrt{\gamma_0/\rho_0}, \qquad 
\beta = \frac{c_t^2}{c_\ell^2} \dfrac{\alpha_2 + \gamma_2 + \gamma_1}{3\gamma_0}.
\end{equation} 

To interpret $c_\ell$ and $c_t$, we recall that the equations of linear elastodynamics are $\mu u_{i,jj} + (\lambda + \mu)u_{j,ij} = \rho_0 u_{i,tt}$, where $\lambda$, $\mu$ are the Lam\'e constants.
Hence, for a linear homogenous longitudinal wave $\vec u = u(X,t) \vec i$, we have $(\lambda + 2\mu)u_{XX} = \rho_0 u_{tt}$. 
The changes of amplitude function \eqref{ampl} and variables \eqref{space} then give
$(\lambda + 2\mu)\tilde u_{\tau\tau} = \rho_0 c^2 \tilde u_{\tau \tau}$, showing that 
\begin{equation}
c=c_\ell = \sqrt{(\lambda+2\mu)/\rho_0},
\end{equation}
the speed of the linear longitudinal wave.
Also, we see from \eqref{beta-gamma} that 
\begin{equation}
\gamma_0 = \gamma_{00} = 2 \left.\left(\dfrac{\partial W}{\partial I_1} + \dfrac{\partial W}{\partial I_2} \right)\right|_{I_1=I_2=3, J=1},
\end{equation}
is the infinitesimal shear modulus, i.e. $\mu$, the second Lam\'e constant, showing that 
\begin{equation}
c_t = \sqrt{\mu/\rho_0},
\end{equation}
the speed of the linear transverse wave.

Eqs. \eqref{eq0a}-\eqref{eq0b} are of the same nature as the ones obtained by Zabolotskaya \cite{ZZ}.  
Zabolotskaya goes on to integrate \eqref{eq0b} with respect to $\tau$ and then to neglect the arbitrary integration function $\phi(\chi, \eta)$ that arises (denoted the \emph{static} part of the displacement), which is justified under certain further assumptions. 
However, that step is not necessary. Indeed, we may first differentiate \eqref{eq0a} with respect to $\tau$ and \eqref{eq0b} with respect to $\eta$, to eliminate $\tilde v_{\eta \tau \tau}$.
Then we introduce the new (dimensionless) variable $U=\tilde{u}_\tau$ to obtain 
\begin{equation} \label{eq0bis}
\left(U_\chi + 3\beta U U_\tau \right)_\tau = \tfrac{1}{2} U_{\eta \eta}.
\end{equation}

Equation~\eqref{eq0bis} is a KZK-type equation in two-dimensional space (named after Khokhlov--Zabolotskaya--Kuznetsov), which describes the propagation and diffraction of directional sound beams \cite{HB}. It reduces to the inviscid Burgers' equation for one-dimensional problems where the longitudinal particle velocity does not vary in the transverse direction ($U_{\eta} = 0$). In this special case, discontinuities in the particle velocity can develop at a given propagation distance from a moving boundary, which is the shock formation distance.

\subsection{Dispersion}

To introduce \emph{dispersion}, we follow the procedure developed in Reference \cite{Destradecompact}. 
Hence we consider that the Cauchy stress is split additively in an elastic part and in a part depending on the first and second Ericksen-Rivlin tensors, as
 \begin{equation} \label{disperso}
 \vec{T}=\vec{T}^e + \nu (\vec{A}_2-\vec{A}_1^2),
 \end{equation}
where $\vec{T}^e = J^{-1} \vec F \vec{S}^e$ is the elastic stress contribution,
\begin{equation}
\vec A_1 = \vec L + \vec L^T, \qquad
\vec{A}_2=\vec{\dot{A}}_1 + \vec{A}_1\vec{L}+\vec{L}^T\vec{A}_1,
\end{equation}
and $\vec L= {\vec{\dot{F}}}\vec{F}^{-1}$ is the velocity gradient (in other words, $\vec A_1 =2\vec D$ where $\vec D$ is the stretching tensor).  
Here the parameter $\nu$ is positive, and its factor $ (\vec{A}_2-\vec{A}_1^2)$ is a gyroscopic term, as discussed in details in References \cite{Destradecompact} and \cite{DestradeSacco}. 
Note also that $\nu/(\rho_0 L^2)$ is non-dimensional \cite{DeSa05}.

For consistency of the expansions, we consider that the parameter $\nu$ is of order $\epsilon$, 
\begin{equation}
\nu = \epsilon \rho_0 L^2\nu_0,
\end{equation}
say, where $\nu_0$ is dimensionless.
Then it suffices to expand $\vec L$, $\vec A_1$, and $\vec{\dot{A}}_1$ to order $\epsilon$
using
\begin{equation}
J \vec L = -\dfrac{\epsilon c_\ell}{L} \begin{bmatrix}
\tilde{u}_{\tau \tau} &0 & 0 \\
0& 0 & 0\\
0 & 0 & 0
\end{bmatrix} = \dfrac{J}2 \vec A_1, \qquad J \vec{\dot{A}}_1 = -\dfrac{2\epsilon c_\ell^2}{L^2} \begin{bmatrix}
\tilde{u}_{\tau \tau \tau} &0 & 0 \\
0& 0 & 0\\
0 & 0 & 0
\end{bmatrix} , 
\end{equation}
ending up with  
\begin{equation}
\nu (\vec A_2 - \vec{A}_1^2)  = -2  \epsilon^2 \rho_0 c_\ell^2 \nu_0 
\begin{bmatrix}
\tilde{u}_{\tau \tau \tau} &0 & 0 \\
0& 0 & 0\\
0 & 0 & 0
\end{bmatrix},
\end{equation}
and neglecting higher-order terms. 

Then we compute the nominal stress $\vec S = J \vec F^{-1}\vec T$ and write the first equation of motion using results above for $\vec S^e$.
After some algebra, we obtain the modified version of equation \eqref{eq0a}  as
\begin{equation} \label{eq0quater}
\left[\left(\frac{c_\ell^2}{c_t^2}-1 \right)\tilde{v}_\eta + 2\frac{c_\ell^2}{c_t^2} \tilde{u}_\chi + 3\frac{ c_\ell^2}{c^2_t}\beta (\tilde{u}_\tau)^2 - 2 \frac{c_\ell^2}{c_t^2} \nu_0 \tilde{u}_{\tau \tau \tau}\right]_\tau = \tilde{u}_{\eta \eta}, 
\end{equation}
while \eqref{eq0b} remains unchanged. 
Using again that equation, and the function $U$, we now end up with 
\begin{equation} \label{eq0bistris}
\left(U_\chi + 3 \beta U U_\tau - \nu_0 U_{\tau \tau \tau} \right)_\tau = \tfrac{1}{2} U_{\eta \eta} ,
\end{equation}
by following the same steps as for the derivation of \eqref{eq0bis}.
The changes of variables from $(\chi, \tau, \eta)$ to $(x,y,t)$ (where the latter is not the original time variable), defined by 
\begin{equation} \label{changevar}
t = -\dfrac{|\beta|^{3/2}}{\sqrt{8\nu_0}} \; \chi, \qquad
x = \dfrac{|\beta|^{1/2}}{\sqrt{2\nu_0}}\;\tau, \qquad
y = \dfrac{|\beta|}{\sqrt{2\nu_0}} \; \eta,
\end{equation}
show that this is indeed the KP-II equation, 
\begin{equation} \label{our-KP-I}
(U_t \mp 6 UU_x + U_{xxx})_x = -U_{yy} ,
\end{equation}
where the minus (or plus) sign is obtained for positive (or negative) $\beta$, respectively. This result is comparable to that obtained in antiplane shear \cite{DestradeGoriely}, except for the nonlinear term.

Finally, reverting \eqref{eq0bistris} to physical quantities \eqref{ampl} and physical coordinates \eqref{space} yields the equation determining $u$ as,
\begin{equation}\label{phys-1}
\left(c_\ell u_{tX} + u_{tt} + 3\dfrac{\beta}{c_\ell}u_tu_{tt} - \dfrac{\nu}{\rho_0 c_\ell^2}u_{tttt}\right)_t = \tfrac{1}{2}c_\ell^2 u_{tYY},
\end{equation}
where we have used the relationship $U = \tilde u_\tau$ in \eqref{eq0bistris}.
Then $v$ is found from 
\begin{equation}\label{phys-2}
v_{tt} + c_\ell u_{tY}=0 ,
\end{equation}
which is a consequence of \eqref{eq0b}.

%%%%%%%%%%%%%%%%%

\section{Plane motions in incompressible solids}
\label{section3}

%%%%%%%%%%%%%%%%%

The equations in the incompressible case may be derived from the above analysis by replacing the constitutive parameter $\alpha$ with an arbitrary Lagrange multiplier $p=p(X,Y,t)$ in the equations \eqref{eqzero} and \eqref{eq}. This Lagrange multiplier is introduced to enforce the internal constraint of isochoricity: $J-1=0$ and, if needed, it is to be determined from initial/boundary conditions.  
All the constitutive information is now concentrated in the parameter $\gamma$, defined in \eqref{beta-gamma}.

However, here the asymptotic analysis is completely different from the compressible case. 
First, we find that for asymptotic consistency, the scaling of the amplitudes and coordinates \eqref{ampl}-\eqref{space} must be replaced by
\begin{subequations} \label{scale-incomp}
\begin{align}
	&u=\epsilon^2 L \tilde{u}, \qquad v=  \epsilon L \tilde{v}, \qquad p = \rho_0 c_t^2 \tilde p, \label{scale-incomp-u} \\
	&\chi=\epsilon^2 \dfrac{X}{L}, \qquad \eta=\epsilon \dfrac{Y}{L}, \qquad \tau= \dfrac{1}{L}(c_t t - X) , \label{scale-incomp-x}
\end{align}
\end{subequations}
where $c_t$ is the speed of the linear shear wave, such that $\rho_0 c_t^2 = \gamma_0=\mu$.
Here, the leading-order displacement component is that along the $Y$ axis, $v$, which is of order $\mathcal{O}(\epsilon)$. From now on, the Lagrange multiplier $p$ replaces the material response function $\alpha$ in the equations of motion \eqref{eq}.

For these scalings, we find 
\begin{equation} \label{Invar-Incomp-Scale}
I_1-3 = \epsilon^2 [ (\tilde v_\tau)^2 + 2 (\tilde v_\eta - \tilde u_\tau)] ,
\qquad
J-1 = \epsilon^2 (\tilde v_\eta - \tilde u_\tau) ,
\end{equation}
at leading order, instead of \eqref{J}. Enforcement of the incompressibility constraint $J=1$ then entails a simplified expression for $I_1-3$. Furthermore, there is a coupling between the longitudinal and transverse displacement components, $\tilde v_\eta = \tilde u_\tau$.

The strain energy and other quantities now depend on only one variable, $I_1$, and we may expand $\gamma$ in powers of $I_1-3$ up to terms in $\epsilon^3$, as
\begin{equation} \label{gamma-inc-1}
\gamma = \gamma_0 + \gamma_1(I_1-3) = \gamma_0 + \epsilon^2 \gamma_1 \tilde v_\tau^2,
\end{equation} 
where $\gamma_0$ and $\gamma_1$ are constants. 
It then follows that
\begin{equation} \label{gamma-inc-2}
\gamma_X = -\epsilon^2 \dfrac{\gamma_1}{L} (\tilde v_\tau^2)_\tau ,
\qquad
\gamma_Y = \epsilon^3\dfrac{\gamma_1}{L}(\tilde v_\tau^2)_\eta,
\end{equation}
so that the governing equations reduce to 
\begin{align}\label{motinc0}
\tilde p_\tau = 0, \qquad \epsilon \tilde p_\eta \tilde v_{\tau\tau} = 0,
\end{align}
at leading order, where we have used the identity $\gamma_0 = \rho_0 c_t^2$.

From \eqref{motinc0}, we deduce the identities $\tilde p_\tau = \tilde p_\eta = 0$. This way, the governing equations \eqref{eq} reduce to 
\begin{align} \label{motinc2}
\tilde p_\chi = \tfrac23 \beta_3 (\tilde v_{\tau}^2)_\tau ,
\qquad
\tfrac{1}{3} \beta_3 [ (\tilde v_\tau^3)_\tau + (\tilde v_\tau^2)_\eta] - \tilde v_{\chi\tau} + \tfrac12\tilde v_{\eta \eta} = 0
\end{align}
at the orders $\epsilon^2$ and $\epsilon^3$, respectively. Here, we have introduced the constant $\beta_3$ such that
\begin{equation} \label{Beta3}
\beta_3 = \dfrac{3}{2}\dfrac{\gamma_1}{\gamma_0} = \dfrac{3}{\mu}\left(\dfrac{\partial^2 W}{\partial I_1^2} + 2 \dfrac{\partial^2 W}{\partial I_1 \partial I_2} + \dfrac{\partial^2 W}{\partial I_2^2}\right)_{I_1=I_2=3} ,
\end{equation}
recalling that $\gamma_1 = \partial \gamma/\partial I_1$ when $I_1=I_2=3$.
Now, in \eqref{motinc2}, differentiation of the first identity with respect to $\eta$ and of the second one with respect to $\tau$ yields
\begin{equation}\label{wavcubic}
(V_{\chi} - \beta_3 V^2 V_\tau)_{\tau} = \tfrac12 V_{\eta \eta} ,
\end{equation}
where we have introduced $V = \tilde v_\tau$. To reach this result, we have used the fact that $\tilde p$ is a function of $\chi$ only.

Equation~\eqref{wavcubic} is a spatially two-dimensional KZK-type equation with cubic nonlinearity, which is used to describe the propagation and diffraction of directional sound beams in incompressible solids \cite{Z}. Here too, shock waves can form in one-dimensional problems where the shearing velocity does not vary in the transverse direction ($V_{\eta} = 0$). Numerical simulations that illustrate the diffraction of the sound beam, the formation of shear shock waves and the generation of higher harmonics can be found in Ref.~\cite{FV}.

The coefficient of cubic nonlinearity above is the same as that introduced by  Zabolotskaya and collaborators \cite{Z,Wochner08}.
In the context of weakly nonlinear elasticity, it requires an expansion of the strain energy density up to the fourth order, as \cite{Z, DestradeOgden}
\begin{equation}
W = \mu \; \text{tr}(\vec E^2) + (A/3)\; \text{tr}(\vec E^3) + D \; \text{tr}(\vec E^2)^2,
\end{equation}
where $\vec E = (\vec F^T \vec F - \vec I)/2$ is the Green-Lagrange strain tensor, and $A$, $D$ are the third- and fourth-order nonlinear Landau constants, respectively.
Then we find that \cite{DPS}
\begin{equation} 
\beta_3 = \dfrac{3}{2}\left(1+\dfrac{A/2+D}{\mu}\right).
\end{equation}

Adding a dispersive term as in the compressible case requires also a different scaling for the parameter $\nu$, which must now be of order $\epsilon^2$, 
\begin{equation} \label{disp-scale}
\nu = \epsilon^2 \rho_0 L^2 \nu_0,
\end{equation}
see \cite{DestradeGoriely} for the justification. 
We then find that
\begin{equation} \label{disp-incomp}
\nu (\vec A_2 - \vec{A}_1^2)  = -  \epsilon^3 \rho_0 c_t^2 \nu_0 
\begin{bmatrix}
0 & \tilde{v}_{\tau \tau \tau}  & 0 \\
\tilde{v}_{\tau \tau \tau}& 0 & 0\\
0 & 0 & 0
\end{bmatrix},
\end{equation}
Introducing this term into the governing equations and following the above steps yields the modified wave equation
\begin{equation} \label{wav-inc}
	(V_{\chi} - \beta_3 V^2 V_\tau - \tfrac12 \nu_0 V_{\tau\tau\tau})_{\tau} = \tfrac12 V_{\eta \eta}
\end{equation}
instead of \eqref{wavcubic}.

The changes of variables from $(\chi, \tau, \eta)$ to $(x,y,t)$ defined by 
\begin{equation} \label{changevar-inc}
t = -\dfrac{|\beta_3|^{3/2}}{6\sqrt{3\nu_0}} \; \chi, \qquad
x = \dfrac{|\beta_3|^{1/2}}{\sqrt{3 \nu_0}}\;\tau, \qquad
y = \dfrac{|\beta_3|}{3\sqrt{\nu_0}} \; \eta,
\end{equation}
show that this is indeed the KP-II equation with cubic nonlinearity, 
\begin{equation} \label{KP-inc}
(V_t \pm 6 V^2V_x + V_{xxx})_x = -V_{yy} ,
\end{equation}
where the plus (or minus) sign is obtained for positive (or negative) $\beta_3$, respectively. For an alternative derivation of the above result, the interested reader is referred to the Appendix \ref{StreamFunc}.

In a similar fashion to the compressible case \eqref{phys-1}-\eqref{phys-2}, the equation \eqref{wav-inc} with $V = \tilde v_\tau$ can then be rewritten in terms of physical variables and coordinates, see \eqref{scale-incomp}. Doing so, we find
\begin{equation}
\left(c_t v_{tX} + v_{tt} - \dfrac{\beta_3}{c_t^2}v_{t}^2v_{tt} - \dfrac{\nu}{2\rho_0 c_t^2}v_{tttt}\right)_{t} = \tfrac{1}{2}c_t^2v_{tYY} ,
\end{equation}
and then $u$ is found from $\tilde v_\eta = \tilde u_\tau$, i.e.,
\begin{equation}
u_{tt} - c_t v_{tY} = 0 ,
\end{equation}
which is a consequence of incompressibility \eqref{Invar-Incomp-Scale}.

%%%%%%%%%%%%%%%%%

\section{Examples}
\label{section4}

%%%%%%%%%%%%%%%%%

We presented a consistent asymptotic derivation of the KP equation \eqref{eq0bistris} in plane elasticity and of its incompressible counterpart \eqref{wav-inc}. 
Using a suitable scaling transformation, these equations may be recast in \emph{canonical forms}, leading to the KP-II equations \eqref{our-KP-I}-\eqref{KP-inc}, which we recall here:  
\begin{align} 
& (U_t \mp 6 UU_x + U_{xxx})_x = -U_{yy} , \label{KP-I}\\
& (V_t \pm 6 V^2V_x + V_{xxx})_x = -V_{yy} \label{KP-Incompr} .
\end{align}
The list of  the remarkable mathematical properties of these equations is too long to review here and relevant references may be found in the papers \cite{Kodama, Wafo,Webb}. A brief overview of the properties of the KP-II equation can be found in the book by Ablowitz and Clarkson \cite{AbloClark}. Note in passing that specific solutions of the KP equation \eqref{KP-I} can be deduced from solutions of the modified KP equation \eqref{KP-Incompr} by means of the Miura transformation \cite{DK}.

%What is interesting is that we have used a rigorous asymptotic 
%procedure and for the dispersive terms a constitutive model which is rooted in the class of simple materials.

\subsection{Travelling waves}

One remarkable property of the KP-II equation is the existence of travelling wave solutions (line solitons). These particular waveforms travel at an angle to the $y$-axis with a constant speed and no distorsion. For the quadratic KP equation \eqref{KP-I} describing compression waves, such a solution takes the form
\begin{equation} \label{sech1}
U(t,x,y)= \mp 2 \kappa^2 \operatorname{sech}^2 [\kappa (x + \theta y - \upsilon t + x_0)] ,
\end{equation}
where the wave speed is given by $\upsilon= 4\kappa^2 + \theta^2$, and $\operatorname{sech}$ is the hyperbolic secant function. In contrast, for the incompressible case \eqref{KP-Incompr} we obtain the solution
\begin{equation} \label{sech2}
V(t,x,y)= \pm \kappa \operatorname{sech} [\kappa (x + \theta y - \upsilon t + x_0)] ,
\end{equation}
with the wave speed $\upsilon= \kappa^2 + \theta^2$. While the wave velocity is a linear function of the wave amplitude in the former case \eqref{sech1}, a quadratic function is obtained in the latter one \eqref{sech2}. Illustrations are provided in Figure \ref{fig:Waves} where the above waveforms are represented for $\kappa=1$ and $\theta =0$ (Section \ref{sectionNum}).

Let us revisit these results by highlighting a connection between the  KP equations and the Boussinesq equations. Introducing the new variables
\begin{equation} \label{d0}
\mathfrak{t}= y, \quad \mathfrak{s}=x - \upsilon t,
\end{equation}
and considering the reductions
\begin{equation}
U(t,x,y)=U(\mathfrak{t}, \mathfrak{s}), \qquad
V(t,x,y)=V(\mathfrak{t}, \mathfrak{s}),
\end{equation}
for \eqref{KP-I} and \eqref{KP-Incompr}, the canonical forms of the KP equation in the compressible and incompressible cases, we arrive at
\begin{align} \label{d1}
 &  U_{\mathfrak{t} \mathfrak{t}} - \upsilon U_{\mathfrak{s} \mathfrak{s}} \mp 3 (U^2)_{\mathfrak{s} \mathfrak{s}} + U_{\mathfrak{s} \mathfrak{s} \mathfrak{s} \mathfrak{s}} = 0 ,\\
 & V_{\mathfrak{t} \mathfrak{t}} - \upsilon V_{\mathfrak{s} \mathfrak{s}} \pm 2 (V^3)_{\mathfrak{s} \mathfrak{s}} + V_{\mathfrak{s} \mathfrak{s} \mathfrak{s} \mathfrak{s}} = 0 .
\end{align}
The former is a Boussinesq equation \cite{CK}, whereas the latter is a modified Boussinesq equation with cubic nonlinearity. The travelling wave solutions \eqref{sech1}-\eqref{sech2} can be obtained from the above partial differential equations as well.

%\vskip1cm
%\color{red}
%Michel, I think that we have to be careful on the sign of the dispersive part in the incompressible case. Then once we are sure I think it suffices to to derive the displacement fields an to plot these quantities. What do you think. If Benjamin wish to contribute maybe a good idea is to use these solutions has initial data for the numerical solution on the whole line so you see what is going on. 
%\color{black}

\subsection{Numerical results}\label{sectionNum}

The numerical resolution of the KP equation has been approached based on a Fourier spectral method by Klein and coauthors \cite{Klein}, which is adapted from Trefethen's code for the KdV equation{\,---\,}a similar algorithm was implemented in the References \cite{Grava, Ablowitz}. The method is based on the use of a two-dimensional Fourier transform in space and of an integrating factor, so that the KP equation \eqref{KP-I} can be rewritten as a differential equation in time
\begin{equation}
	\big(\text{e}^{t \widehat{\mathcal L}} \widehat{U}\big)_t = \pm 3 \text i k_x \text{e}^{t \widehat{\mathcal L}} \widehat{U^2} , \quad \widehat{\mathcal L} = \frac{\text i k_y^2}{k_x - 0 \text{i}} - \text i k_x^3 ,
	\label{Spectral}
\end{equation}
where $\widehat{U}$ denotes the spatial Fourier transform of $U$, `$\text{i}$' is the imaginary unit, and $(k_x, k_y)$ are the wavenumbers (i.e., the spatial coordinates in Fourier space). The spatial Fourier transforms are evaluated by means of \textsc{Matlab}'s Fast Fourier Transform algorithm \texttt{fft2}, and integration in time is performed using a fourth-order Runge-Kutta scheme. The same method can be used for the numerical resolution of the modified KP equation \eqref{KP-Incompr} up to minor changes. Most notably, the Fourier transform of $U^2$ in Eq.~\eqref{Spectral} needs to be replaced by the Fourier transform of $V^3$, and coefficients need to be adjusted.

\begin{figure}
	\centering
	
	\includegraphics[width=0.7\textwidth]{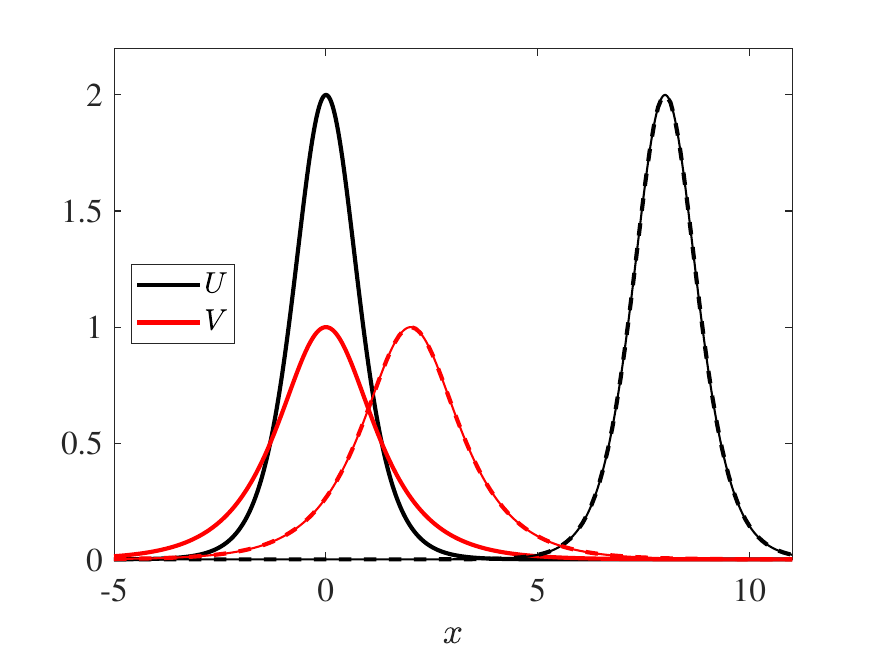}
	
	\caption{Solitary waves. Thick solid lines: initial conditions \eqref{Initial} for the nonlinear wave equations \eqref{KP-I} and \eqref{KP-Incompr}. Thin solid lines: exact solution \eqref{sech1}-\eqref{sech2} at $t=2$. Dashed lines: waveforms obtained along the line $y=0$ at $t=2$ using a Fourier spectral method. Integration was performed for $256 \times 256$ points $(x,y)$ in $[-4\pi , 4\pi]^2$, and a time step of $10^{-4}$. \label{fig:Waves}}
\end{figure}

For the KP equations in canonical form, we select the initial conditions
\begin{equation}
	U(0,x,y) = 2\, \text{sech}^2(x) , \quad 
	V(0,x,y) = \text{sech}(x) ,
	\label{Initial}
\end{equation}
which correspond to the initial waveform of a solitary wave solution to the nonlinear wave equations \eqref{KP-I} and \eqref{KP-Incompr} with the appropriate signs, see Eqs. \eqref{sech1} and \eqref{sech2}. These waves propagate along increasing $x$ with an invariant waveform at a constant speed equal to 4 and 1, respectively. The initial conditions \eqref{Initial} correspond to the special values $\kappa=1$ and $\theta=0$ in both cases.

We have represented these initial conditions in Figure \ref{fig:Waves}, as well as the waveforms at $t\simeq 2$ after sufficient iterations of the spectral method \eqref{Spectral}. Here, the results were obtained by using $256 \times 256$ points for $(x,y)$ within the domain $[-4\pi , 4\pi]^2$, while the iterations in time were performed with a time step of $10^{-4}$. The null imaginary number $0\text{i}$ in the denominator of Eq.~\eqref{Spectral} is replaced by a small imaginary number near $10^{-16} \text{i}$ to avoid singularity. As expected, we observe that the solitary waves computed numerically propagate with a constant shape at a speed of $4$ and $1$ units, respectively. Thus, the spectral method provides an accurate computational solution to the initial-value problems \eqref{Initial}.

\begin{figure}
	\centering

	\includegraphics[width=\textwidth]{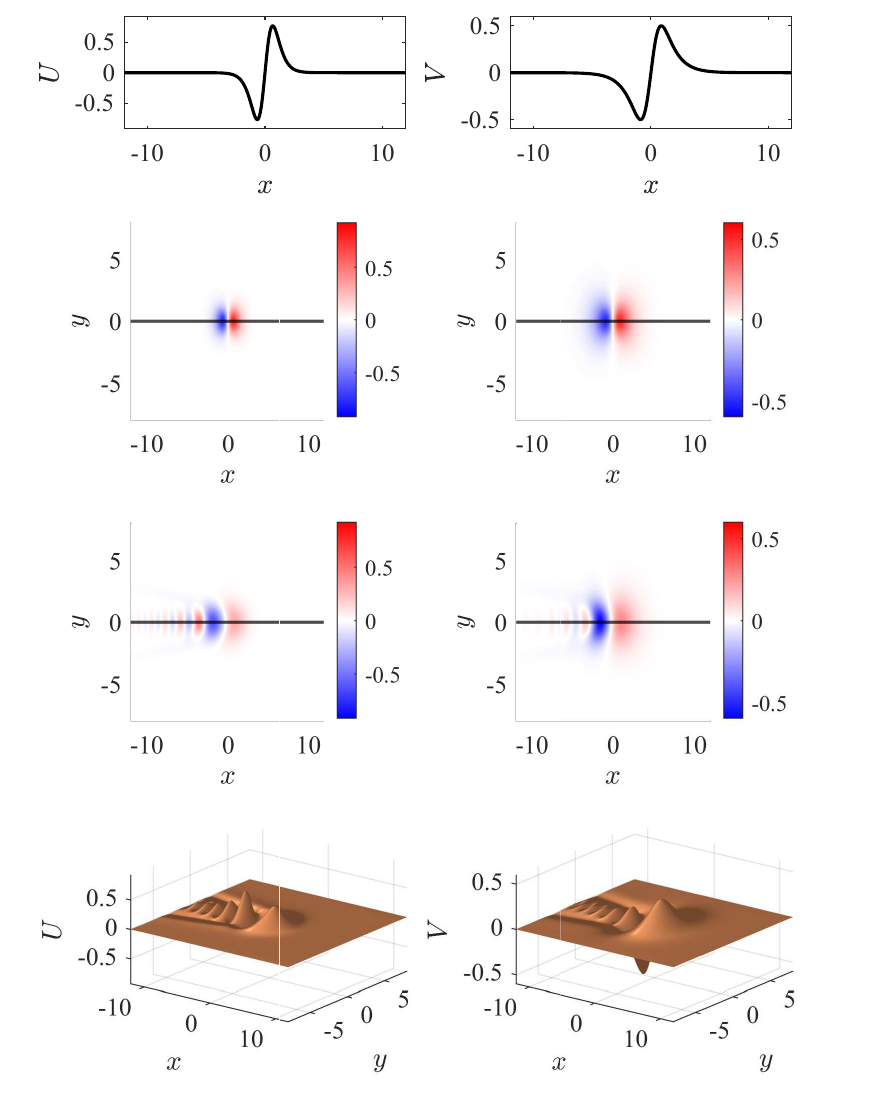}
	
	\caption{Initial-value problems \eqref{Initial2}, where $t=0$ in the first and second rows of figures. The nonlinear wave equations \eqref{KP-I} and \eqref{KP-Incompr} are solved in the left and right column, respectively. The third and fourth rows of figures show the solutions obtained numerically at $t=0.2$. Integration was performed for $2048 \times 1024$ points $(x,y)$ in $[-16\pi , 16\pi] \times [-8\pi , 8\pi]$, and a time step of $0.5\times 10^{-4}$. \label{fig:Peaks}}
\end{figure}

Using the same algorithm, we now select the initial conditions
\begin{equation}
	U(0,x,y) = -\frac{\partial}{\partial x} \text{sech}^2(r) , \quad 
	V(0,x,y) = -\frac{\partial}{\partial x} \text{sech}(r) ,
	\label{Initial2}
\end{equation}
where $r = \sqrt{x^2 + y^2}$. These functions correspond to those in Eq. \eqref{Initial} with a smooth envelope, see also Grava et al. \cite{Grava}. The results shown in Figure \ref{fig:Peaks} were obtained by using $2048 \times 1024$ points for $(x,y)$ within the domain $[-16\pi , 16\pi] \times [-8\pi , 8\pi]$, while the iterations in time were performed with a time step of $0.5\times 10^{-4}$ up to the final time, $t\simeq 0.2$.

Qualitatively, the numerical solution obtained for Eq.~\eqref{KP-I} (left column in Figure \ref{fig:Peaks}) is comparable to the one obtained for Eq.~\eqref{KP-Incompr} (right column in the Figure). In the bottom row of Figure \ref{fig:Peaks}, the multiple oscillations that have formed along the $x$-axis result from the interaction between nonlinearity and wave dispersion (left-hand side of Eqs.~\eqref{KP-I}-\eqref{KP-Incompr}), in a similar fashion to the formation of oscillations in the KdV equation \cite{BK}. The evolution of the wave fields along the $y$-axis is caused by the diffraction term in the right-hand side of Eqs.~\eqref{KP-I}-\eqref{KP-Incompr}, in a similar fashion to the diffraction of sound beams in the KZK equation \cite{FV}. For further physical interpretations, it is essential to remember that the coordinates $(t,x,y)$ defined in Eqs.~\eqref{changevar}-\eqref{changevar-inc} are scaled versions of a propagation distance ($\propto X$), a retarded time variable ($\propto [t-X/c]$), and a transverse spatial coordinate ($\propto Y$), respectively. Practical implications of the observed features and of other properties shall become the scope of future research.

%%%%%%%%%%%%%%%%%

\section{Conclusion}

We have shown that the KP equation can be derived within the context of plane compressible elasticity with dispersion, based on a suitable re-scaling of the equations of motion. Following similar steps, a modified KP equation with cubic nonlinearity is obtained in the incompressible case, with some technical differences compared to the compressible case. These partial differential equations describe nonlinear wave motion in solids with a preferred propagation direction.

Solutions to these partial differential equations include solitary waves and other nonlinear waves. In fact, numerical simulations based on a Fourier spectral method show that solutions to the KP equation and to its cubic counterpart are essentially similar (at least qualitatively). In line with other works from the literature \cite{Ablowitz,Grava}, a detailed analysis of the dispersive shock wave solutions might become the scope of future works.

\appendix
\section{Stream function for the incompressible case}\label{StreamFunc}

We present an alternative derivation for the incompressible case. The main motivation for this Appendix is to address the slight mismatch of the asymptotic orders in Section \ref{section3}, where the first equation of motion \eqref{eqa} is solved up to order $\epsilon^2$ whereas the second one \eqref{eqb} is solved up to order $\epsilon^3$. Here, we assume that the scaling of the amplitude is of order $\epsilon$ for both $u$ and $v$, in contrast to \eqref{ampl} and \eqref{scale-incomp-u}:
\begin{equation} \label{scale-stream1}
u = \epsilon L \tilde u, \qquad 
v = \epsilon L \tilde v. 
\end{equation}

We may then introduce a `stream' function $\psi=\psi(X,Y,t)$, defined by $\psi_{X} = -v$, $\psi_Y=u$, with non-dimensional measure $\tilde \psi$ defined by 
\begin{equation} \label{scale-stream2}
\psi=\epsilon L^2 \tilde{\psi}.
\end{equation} 
It follows that the invariants $I_1$ and $J$ in \eqref{I1-J} now read
\begin{equation}
I_1-3=\epsilon^2L^4(\tilde\psi_{XX}^2 + \tilde\psi_{YY}^2 + 2\tilde\psi_{XY}^2),
\quad
J-1=\epsilon^2L^4(\tilde\psi_{XX}\tilde\psi_{YY} - \tilde\psi_{XY}^2).
\label{I1-J-inc}
\end{equation}
Similarly, the equations of motion \eqref{eq} are rewritten in terms of $\tilde\psi$ as
\begin{equation} \label{eq-inc}
\begin{aligned}
& p_{X} (1 - \epsilon L^2\tilde\psi_{XY}) + p_{Y} \epsilon L^2\tilde\psi_{XX} + \gamma\epsilon L^2(\tilde\psi_{XXY} + \tilde\psi_{YYY}) \\
&\qquad\qquad\qquad\qquad + \gamma_{X}(1+ \epsilon L^2\tilde\psi_{XY}) + \gamma_{Y} \epsilon L^2\tilde\psi_{YY} = \rho_0  \epsilon L^2\tilde\psi_{Ytt},  \\[8pt]
& {-} p_{X} \epsilon L^2\tilde\psi_{YY} + p_{Y}(1 + \epsilon L^2\tilde\psi_{XY}) - \gamma \epsilon L^2 (\tilde\psi_{XXX}  + \tilde\psi_{XYY}) \\
&\qquad\qquad\qquad\qquad - \gamma_{X}\epsilon L^2\tilde\psi_{XX} + \gamma_{Y}(1-\epsilon L^2\tilde\psi_{XY}) =- \rho_0   \epsilon L^2\tilde\psi_{Xtt}.
\end{aligned}
\end{equation}

Next, we find that the scaling of the space and time coordinates is not of the form \eqref{space}; it should now be of the same form as in \eqref{scale-incomp-x}. Then the relevant derivatives of $\tilde\psi$ up to order $\epsilon^3$ follow as
\begin{equation}
\begin{aligned}
&\tilde\psi_{XX} = \dfrac{1}{L^2}\tilde\psi_{\tau\tau} - 2\dfrac{\epsilon^2}{L^2}\tilde\psi_{\chi\tau}, 
\quad
\tilde\psi_{XY} = -\dfrac{\epsilon}{L^2}\tilde\psi_{\eta\tau} + \dfrac{\epsilon^3}{L^2}\tilde\psi_{\chi\eta}, 
\quad
\tilde\psi_{YY} = \dfrac{\epsilon^2}{L^2}\tilde\psi_{\eta\eta}, \\[4pt]
&\tilde\psi_{XXX} = -\dfrac{1}{L^3}\tilde\psi_{\tau\tau\tau} + 3\dfrac{\epsilon^2}{L^3}\tilde\psi_{\chi\tau\tau}, 
\qquad
\tilde\psi_{XXY} = \dfrac{\epsilon}{L^3}\tilde\psi_{\eta\tau\tau} -2 \dfrac{\epsilon^3}{L^3}\tilde\psi_{\chi\eta\tau},  \\[4pt]
& \tilde\psi_{XYY} = -\dfrac{\epsilon^2}{L^3}\tilde\psi_{\eta\eta\tau}, 
\qquad
\tilde\psi_{YYY} = \dfrac{\epsilon^3}{L^3}\tilde\psi_{\eta\eta\eta}, \\[4pt]
& \tilde\psi_{Xtt} = -\dfrac{c_t^2}{L^3}\tilde\psi_{\tau\tau\tau} + \epsilon^2\dfrac{c_t^2}{L^3}\tilde\psi_{\chi\tau\tau}, 
\qquad
\tilde\psi_{Ytt} = \epsilon\dfrac{c_t^2}{L^3}\tilde\psi_{\eta\tau\tau}.
\end{aligned}
\end{equation}
As a result, Eq.~\eqref{I1-J-inc} expanded to order $\epsilon^3$ now yields $I_1-3 = \epsilon^2 \tilde\psi_{\tau\tau}^2$, and the incompressibility constraint $J-1=0$ is satisfied.

Next, we expand $\gamma$ in powers of $I_1-3$ up to terms in $\epsilon^3$. This way, we recover Eqs.~\eqref{gamma-inc-1}-\eqref{gamma-inc-2} with $\tilde v_{\tau}^2$ replaced by $\tilde \psi_{\tau\tau}^2$.
Up to order $\epsilon^3$, the governing equations \eqref{eq-inc} reduce to 
\begin{equation} \label{eq-inc2}
\begin{aligned} 
& (1 + \epsilon^2 \tilde\psi_{\eta\tau})p_X  +  (\epsilon \tilde\psi_{\tau\tau} - 2\epsilon^3 \tilde\psi_{\chi\tau})p_Y =  \epsilon^2 \dfrac{\gamma_1}{L}(\tilde\psi_{\tau\tau}^2)_\tau, \\
& {-} \epsilon^3 \tilde\psi_{\eta\eta}p_X + (1 - \epsilon^2 \tilde\psi_{\eta\tau})p_Y = \dfrac{\epsilon^3}{L}\{\gamma_0(2\tilde\psi_{\chi\tau\tau} - \tilde\psi_{\eta\eta\tau}) - \gamma_1[(\tilde\psi_{\tau\tau}^3)_\tau + (\tilde\psi_{\tau\tau}^2)_\eta]\}.
\end{aligned}
\end{equation}

We may now solve this system for $p_X$ and $p_Y$, as
\begin{equation} \label{solvep}
\begin{aligned}
& p_X = \dfrac{\epsilon^2}{L} \gamma_1(\tilde\psi_{\tau\tau}^2)_\tau,  \\
& p_Y  = \dfrac{\epsilon^3}{L}\{\gamma_0(2\tilde\psi_{\chi\tau\tau} - \tilde\psi_{\eta\eta\tau}) - \gamma_1[(\tilde\psi_{\tau\tau}^3)_\tau + (\tilde\psi_{\tau\tau}^2)_\eta]\} ,
\end{aligned}
\end{equation}
where higher-order terms have been neglected. 
Finally, writing that the cross derivatives of $p$ must be equal, $p_{YX} = p_{XY}$, we arrive at the following cubic nonlinear equation in  $\Psi=\tilde{\psi}_{\tau \tau}$,
\begin{equation} \label{cubic}
(\Psi_{ \chi} - \beta_3\Psi^2\Psi_{ \tau})_\tau = \tfrac{1}{2}\Psi_{\eta \eta},
\end{equation}
where the coefficient $\beta_3$ was introduced in Eq.~\eqref{Beta3}.
 
Eq.~\eqref{cubic} is a special case of the generalized Zabolotskaya (GZ) system \cite{Pucci0,DPS}. It can also be established from the results in Ref.~\cite{DPS}, where in addition to the in-plane motion described by $u$, $v$, an out-of-plane motion $w(X,Y,t)$ was also considered. 
The end result was a coupled system of nonlinear equations for $\psi$ and $w$. 
Taking $w=0$, as here, in that system, leads to the single equation \eqref{cubic} for $\Psi$, albeit with a sign mistake for the first term of the equation. 

Adding a dispersive term requires the same scaling for the parameter $\nu$ as in Eq.~\eqref{disp-scale}, namely $\nu = \epsilon^2 \rho_0 L^2 \nu_0$. We then find that $\nu (\vec A_2 - \vec{A}_1^2)$ is of the same form as \eqref{disp-incomp} with $\tilde v$ replaced by $\tilde{\psi}_{\tau}$, at leading order. Introducing this term into the governing equations \eqref{eq-inc} and solving for $p_X$ and $p_Y$ now gives
\begin{equation}
\begin{aligned}
& p_X = \dfrac{\epsilon^2}{L^2} \gamma_1(\tilde\psi_{\tau\tau}^2)_\tau, \\
& p_Y  = \dfrac{\epsilon^3}{L}\{\gamma_0(2\tilde\psi_{\chi\tau\tau} - \tilde\psi_{\eta\eta\tau} - \nu_0 \tilde{\psi}_{\tau\tau \tau \tau \tau}) - \gamma_1[(\tilde\psi_{\tau\tau}^3)_\tau + (\tilde\psi_{\tau\tau}^2)_\eta]\},
\end{aligned}
\end{equation}
instead of \eqref{solvep}, and writing $p_{YX}=p_{XY}$ leads to 
\begin{equation}
( \Psi_{ \chi } - \beta_3\Psi^2\Psi_\tau - \tfrac{1}{2}\nu_0 \Psi_{\tau\tau \tau})_\tau = \tfrac{1}{2}\Psi_{\eta \eta}  
\end{equation}
instead of \eqref{cubic}.
This equation is of the exact same form as \eqref{wav-inc}. Likewise, it can be reduced to the canonical form \eqref{KP-inc} of the modified KP-II equation with cubic nonlinearity.

The correspondence between the results in Section \ref{section3} and in the Appendix \ref{StreamFunc} can be established based on the respective scaling assumptions \eqref{scale-incomp-u} and \eqref{scale-stream1}-\eqref{scale-stream2}, together with \eqref{scale-incomp-x}. In fact, according to the definition of the stream function $\psi$, we note that the following quantities are equal,
\begin{equation}
	\begin{aligned}
	\psi_X &= -\epsilon L \tilde \psi_\tau + \epsilon^3 L \tilde \psi_\chi ,\\
	  &= -v = -\epsilon L \tilde v ,
	 \end{aligned}
\end{equation}
where the last equality holds for the derivations in Section \ref{section3} and in the present Appendix. We can therefore readily equate $\Psi = \tilde \psi_{\tau \tau}$ with $V = \tilde v_\tau$ to the same order of approximation, see the notations used in \eqref{wavcubic} and \eqref{wav-inc}.

%\begin{lemma}
%Test Lemma.
%\end{lemma}
%
%
%\section*{Acknowledgments}
%We would like to acknowledge the assistance of volunteers in putting
%together this example manuscript and supplement.

%%%%%%%%%%%%% References biblio %%%%%%%%%%%%%%%
\addcontentsline{toc}{section}{References}

%%------------ MRC BIBLIO
%\section*{\refname}
%\addcontentsline{toc}{section}{References}
\bibliography{references}{}

\end{document}